\documentclass[conference]{IEEEtran}
\IEEEoverridecommandlockouts
\usepackage{cite}
\usepackage{amsmath,amssymb,amsfonts}
\usepackage{pstricks}
\usepackage{graphicx}
\usepackage{textcomp}
\usepackage{xcolor}
\usepackage{multirow}
\usepackage{bbm}
\usepackage{dsfont}
\usepackage{comment}
\usepackage{breqn}
\usepackage{amsmath}
\usepackage{caption}
\usepackage{subcaption}
\usepackage{multicol}
\newcommand{\nonl}{\renewcommand{\nl}{\let\nl\oldnl}}
\def\BibTeX{{\rm B\kern-.05em{\sc i\kern-.025em b}\kern-.08em
    T\kern-.1667em\lower.7ex\hbox{E}\kern-.125emX}}
    
\usepackage{eso-pic}
\newcommand\AtPageUpperMyright[1]{\AtPageUpperLeft{%
 \put(\LenToUnit{0.5\paperwidth},\LenToUnit{-1cm}){%
     \parbox{0.5\textwidth}{\raggedleft\fontsize{9}{11}\selectfont #1}}%
 }}%
\newcommand{\conf}[1]{%
\AddToShipoutPictureBG*{%
\AtPageUpperMyright{#1}
}
}

\begin{document}



\title{Programmable FPGA-based Memory Controller}


\author{
    \IEEEauthorblockN{Sasindu Wijeratne\IEEEauthorrefmark{1}, Sanket Pattnaik\IEEEauthorrefmark{1}, Zhiyu Chen\IEEEauthorrefmark{1},  Rajgopal Kannan\IEEEauthorrefmark{2}, Viktor Prasanna\IEEEauthorrefmark{1}}
    \IEEEauthorblockA{\IEEEauthorrefmark{1}Department of Electrical and Computer Engineering, University of Southern California, Los Angeles, USA}
    \IEEEauthorblockA{\IEEEauthorrefmark{2}US Army Research Lab, Los Angeles, USA}
    Email: \{kangaram, spattnai, zhiyuc\}@usc.edu, rajgopal.kannan.civ@mail.mil, prasanna@usc.edu
}

\maketitle

\conf{IEEE 28th International Conference on Hot Interconnects, 2021}

\begin{abstract}
Even with generational improvements in DRAM technology, memory access latency still remains the major bottleneck for application accelerators, primarily due to limitations in memory interface IPs which cannot fully account for variations in target applications, the algorithms used, and accelerator architectures. Since developing memory controllers for different applications is time-consuming, this paper introduces a modular and programmable memory controller that can be configured for different target applications on available hardware resources. The proposed memory controller efficiently supports cache-line accesses along with bulk memory transfers. The user can configure the controller depending on the available logic resources on the FPGA, memory access pattern, and external memory specifications. The modular design supports various memory access optimization techniques including, request scheduling, internal caching, and direct memory access. These techniques contribute to reducing the overall latency while maintaining high sustained bandwidth. We implement the system on a state-of-the-art FPGA and evaluate its performance using two widely studied domains: graph analytics and deep learning workloads. We show improved overall memory access time up to 58\% on CNN and GCN workloads compared with commercial memory controller IPs.
\end{abstract}


\section{Introduction}
Traditional CPU designs are challenged in meeting the computational requirement of big data analytics applications. Accelerators for such applications on Field Programmable Logic Arrays (FPGAs) have become an attractive solution due to their massive parallelism, low power consumption, and cost-efficiency \cite{9049213}\cite{8445087}. Unfortunately, effective external DRAM memory bandwidth and access latency have become the bottleneck in such accelerators \cite{electronics10040438}\cite{10.1145/2534845}. Generational improvements in external memory have primarily targeted storage capacity and bandwidth.  On the other hand,  external memory access latency has seen limited improvements over the past few decades. %
Accessing main memory can cost multiple clock cycles depending on the memory access pattern and can lead up to 70\% idle time on the FPGA accelerator \cite{6903697}\cite{6339373}.

There have been several techniques proposed to overcome the access latency \cite{10.1145/977091.977115}\cite{volos2016an}. Caches \cite{5695314} are very productive in this regard if the required data fits in the cache and the data has spatial and temporal locality\cite{10.1145/106972.106981}. The ongoing approach for solving long memory access delays is to use onboard Block RAM (BRAM) in the FPGA as a data cache and facilitate data retrieval. An alternative solution is to look at multiple memory requests as \textit{bulk memory transfers}\cite{10.1145/3020078.3021743}.

Although FPGA-based designs use custom logic tailored for individual applications, interfaces like main memory are still comparable to traditional computing platforms like CPUs and GPUs. As such, even with a custom memory hierarchy, accessing lower-level memory structures prevents low latency computations. Also, applications with irregular memory accesses render caches ineffective \cite{10.1145/2089142.2089151}. Therefore, the performance boost with custom parallel FPGA hardware is limited by how fast the data can be transferred between the accelerator and main memory.

The heterogeneity of FPGA technology has enabled applications with different physical and hardware requirements to use FPGAs. The resource consumption of various accelerators can vary drastically. A parameterized memory controller is more favorable because the memory controller can be remodelled depending on the targeted application. Also, using excessive hardware resources on the memory controller can have no impact due to the throughput saturation by physical constraints (i.e., DRAM technology, PCB routing, memory access protocols, etc.).

In this paper, we propose a modular and programmable memory controller. It is used as an enhanced memory controller to reduce total memory access time. Our work can integrate with different FPGA accelerators that execute numerous complex algorithms on various FPGA platforms. The proposed design has a set of re-configurable parameters. Those parameters are configured during the synthesis time depending on the functional specification of the accelerator, FPGA platform, and available hardware resources. In the results section, we demonstrate the usability and impact of the proposed system by integrating it with several types of accelerators that target different algorithms.
 
The multi-faceted approach helps towards obtaining low latency as well as high bandwidth. The paper also motivates to cumulate memory requests over time (i.e., forming a batch) and schedule them while exploring their locality in space before accessing the memory. The parallelism of FPGAs is leveraged to reorder the requests in a batch faster and more efficiently. The batch-wise reordering improves row buffer hits within the DRAM \cite{6237036}. It minimizes the total row switching overhead and ensures the lowest possible memory latency for particular batch size. In the meantime, the memory controller also aims to fulfill the traditional memory hierarchy by introducing an inbuilt cache that serves to quickly service requests from the accelerator's Processing Elements (PE). Depending on the external memory configuration and data traffic to the FPGA, the memory controller can be modified to support batches of varying sizes. The programmability of our design allows the controller to be customized based on application requirements and performance objectives. It also facilitates the creation and modification of accelerator-specific control policies. 


The key contributions of this paper are:
 \begin{itemize}	
     
     \item \textbf{Programmable memory controller}: Our design principles focus on a user-configurable memory controller. Within the context of FPGA-driven applications, programmability allows for design parameters introduced in section II to be modified. The configured memory controller can then deployed depending on application traffic and Memory access patterns.  
        
    \item\textbf{ Modularity}: The modular design allows the core components such as cache, Direct Memory Access (DMA), and memory scheduler to be deployed independently with minor changes to the parameters of our memory controller IP depending on the user and application requirements. 

    \item We evaluate the memory controller on a Xilinx Alveo U250 board and target GCN and CNN workloads. Our work improves overall memory access time up to 58\% compared to commercial memory controllers\cite{xilinxddr}\cite{intelddr}.
     
 \end{itemize}

The rest of the paper is organized as follows, Section II and III focus on defining the reconfigurability of the memory controller and the prior work. Section IV discusses the architecture of the proposed controller design in-depth, while Section V presents the evaluation results.


\begin{figure*}[t] 
\centering
\includegraphics[width=\textwidth]{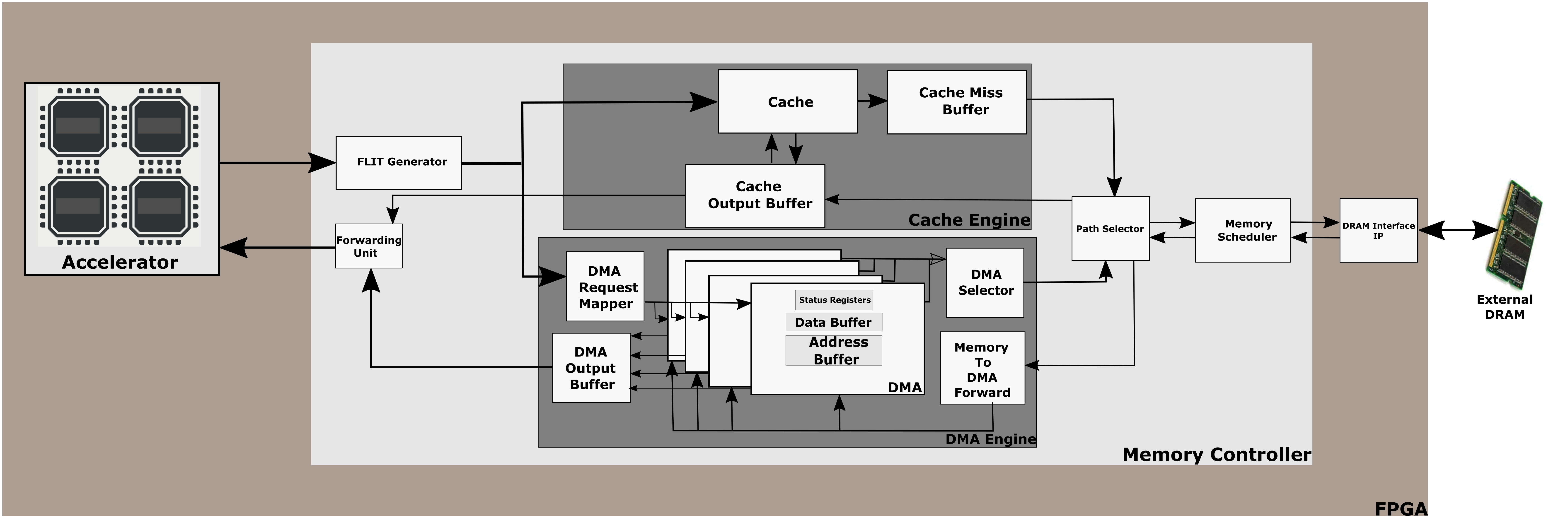}
\caption{Overall Architecture}
\label{overall}
\end{figure*}


\section{Reconfigurable Memory Controller}\label{reconfig_mem}
Our memory controller is a configurable Intellectual Property (IP) block. The configurable parameters of our IP have explicit dependencies on the FPGA platform (PL), available FPGA resources (RS), and functional specifications (SPEC) of the targeted application/accelerator. Few parameters need to be manually tuned (TUNE) to get better performance. Table \ref{re_param} shows a summary of all the configurable parameters in the controller.
\begin{table}[ht!] 
\centering
\caption{Reconfigurable Memory Controller Parameters}
\label{re_param}
\begin{tabular}{|c|c|c|}\hline
\textbf{Parameter} & \textbf{Typical Value Range} & \textbf{Dependency} \\\hline\hline
\multicolumn{3}{ |c| }{\textbf{Overall Design}} \\\hline
Memory interface data width & 64B - 512B & PL \\\hline
Memory interface address width & 20-36 & PL\\\hline
Application I/O data width &	1B – 64B &	SPEC\\\hline
Application address width &	28 - 37	& SPEC\\\hline
Number of PEs & 1 - 128 & SPEC\\\hline
Enable Scheduler &	1 / 0	& SPEC\\\hline
Enable Cacheline &	1 / 0	& SPEC\\\hline
Enable DMA &	1 / 0	& SPEC\\\hline\hline
\multicolumn{3}{ |c| }{\textbf{Direct Memory Access (DMA)}} \\\hline
Maximum transaction size &	256B – 256KB  &	SPEC\\\hline
No. parallel DMA transactions & 1-8  &	SPEC\\
 & & TUNE\\\hline\hline
\multicolumn{3}{ |c| }{\textbf{Memory Scheduler}} \\\hline
Scheduling batch size & 4-128  & TUNE\\\hline
Scheduler timeout & 4 – 40 cycles  & TUNE\\\hline\hline
\multicolumn{3}{ |c| }{\textbf{Cache}} \\\hline\hline
Cache line width & 256-1024  &	SPEC\\
 & & PL\\
 & & RS\\\hline
Num. cache lines & 256 – 16K  & SPEC\\
 & & RS\\\hline
Associativity (DoSA) & 1 - 16  & TUNE\\
 & & RS\\\hline
 
\end{tabular}
\vspace{-0.2cm}
\end{table}

\subsubsection{Cache Engine} \label{cache_sub}
The users can configure cache line width, cache line count, and associativity in the cache.
 
\subsubsection{DMA Engine} \label{dma_sub}

The DMA engine supports simultaneous bulk transfers. The design parametrizes the number of parallel DMA transfers base on the available resources. The users can also update the DMA buffer size depending on the average bulk transfer size of the connected accelerator.

 \subsubsection{Scheduler} \label{scheduler_sub}

The objective of a reconfigurable memory scheduler is to change the size of a batch and scheduler timeout based on the targeted application's latency requirement and available resources in the FPGA. The batch size determines the maximum number of requests the scheduler can process/rearrange in parallel. The timeout decides the maximum duration spent on batch formation using incoming memory requests.

\section{Related Work}
S. Aananthakrishnan et al.\cite{ DBLP:journals/corr/abs-2010-06277} have proposed a memory-optimized large-scale graph processor on ASICs. This paper emphasizes the importance of parallel support for bulk transfers and cache transfers for graph workloads. Our design differs from this work due to the modularity and configurability of our architecture. Also, our memory controller is not bound to any specific accelerator and can use as a generic component with various FPGA-based accelerators after configuration.

X. Ma et al. \cite{10.1145/3020078.3021743} developed a lightweight graph processing engine on FPGA focusing on optimizing the external memory access of each thread. Each of the PEs in the accelerator has its separate private transactional memory supporting bulk transactions. Since a single graph application can have a mixture of regular and irregular data access patterns \cite{osti_10125453}, replacing the cache entirely can negatively impact the performance. In our approach, we support multiple bulk transfers in parallel while maintaining a separate cache pipeline.

Z. Shao et al. \cite{10.1145/3289602.3293900} propose a multi-level buffering technique to reduce the total memory access latency. This work stores all the incoming data on global buffers first and then forwards to the PEs. Such an approach can led to excessive use of global buffering, which can increase the hardware complexity of the implementation. In our work, only the re-usable data structures are globally cached using the internal SRAMs.  Hence it significantly decreases the size of the cache/buffers.

Bojnordi and Ipek\cite{10.1145/2534845} describe a programmable memory controller design for ASICs. This work is the only programmable memory controller design in literature prior to our work. While this approach explores an Instruction Set Architecture (ISA) for a modular design, we take advantage of the reconfigurability of the FPGA to explore the programmable aspect. \cite{10.1145/2534845} does not focus on supporting DMA transfers separately.

R. Wei et al.\cite{electronics10040438} present an FPGA-based memory scheduler with multiple scheduling policies to support Neural Network acceleration. Their approach predicts an access policy based on recent access patterns. But the proposed deep sequential pipeline in the paper introduces extra latency to the overall access time. In our approach, we use a lightweight memory scheduling policy that explores the locality of the memory requests at a given period while exploring available parallelism. 

G. Csordas et al. \cite{8977899} investigate the importance of re-ordering memory accesses to improve the bandwidth. The paper lacks implementation details of the work. In our paper, we propose a practical implementation for memory re-ordering while exploiting the parallelism of FPGAs. 

DMA-controlled hardware accelerators are commonly used in CNN applications. Typically, the communication performance improves with the number of DMAs at the cost of additional hardware complexity.  T. Chen et al. \cite{10.1145/2654822.2541967} present a hardware accelerator that is equipped with two separate DMACs, one for data input and one for data output. Also, L. Bai et al. \cite{9180842} uses multiple read DMAs to improve communication performance. However, none of these approaches focus on global caching techniques to minimize the overall access time.

Our novel approach involves combining lightweight, modular components to allow maximum flexibility to the user. Our \textit{plug and play} approach seeks the user to take advantage of the reconfigurability of the FPGA for a diverse range of constraints such as resource utilization, platforms, and applications due to a rich set of tunable parameters.
\begin{table}[ht!] 
\centering
\caption{Notation}
\label{notation}
\begin{tabular}{c|c}\hline
\textbf{Notation} & \textbf{Description}
\\\hline\hline
$T_s$    &	Total time spent on process \textit{s}	\\\hline
$L_s$  &	Latency of process \textit{s}	\\\hline
$\mathds{1}(\cdot)$	&	$ \mathds{1}(K \;=\; k) = \begin{cases} 1 & (K == k) \\ 0 & else \\ \end{cases} $	\\\hline

\end{tabular}
\end{table}
\section{Memory Controller Architecture}

\textbf{Notation: } In this paper, we use some notations to explain our idea clearly. Table \ref{notation} summarizes the symbols commonly used in the paper.

We develop a unified reconfigurable memory controller template that can be shared among several PEs in an accelerator. The memory controller can be configured depending on the FPGA platform, available resources, and functional specifications of the PEs. Section II describes more details of the reconfigurability of our memory controller. 

Our proposed architecture supports 2 types of memory transfers:
\begin{enumerate}
  \item \textbf{Cache-line transfers}: Nurtures single memory accesses. Load/store operations on a single requested data with minimum latency.
  \item \textbf{Bulk transfers}: Supports streaming accesses. Load/store operations on all requested data with minimum latency from memory.
\end{enumerate}	

Figure \ref{overall} shows the overall architecture of the memory controller. Our memory controller requires several inputs from PEs to initiate a request. These inputs include PE id, access type, payload size, total request size, memory address, and the payload. The width of PE id ($\propto$ number of PEs), payload size ($\propto$ data width), and memory address width depend on the reconfigurable parameters. Internal modules communicate with each other using Flow Control Units (FLITs). It increases communication efficiency between modules while reducing the latency overheads. Hence, as soon as a request reaches the controller, it is converted to a FLIT using the FLIT generator. FLIT generator sends the FLIT header and the payload separately and reduces routing congestion and payload traffic. After that, FLIT forwards the request to either the cache engine or DMA engine, depending on the access type. Our memory controller has latency overhead ($L_{ctrl\_oh}$) due to FLIT generation and path selection logic. We keep $L_{ctrl\_oh}$  limited to 10 cycles by using efficient FLIT encoding and decoding logic.

The cache engine and DMA engine are responsible for cache-line transfers and bulk transfers, respectively. When cache and DMA try to access the memory simultaneously, the cache-line receives higher priority because DMA transfers take a comparatively longer time than a cache-line transfer. But, after a DMA transfer is enabled, the cache-line is stalled until the active DMA transfer is complete.

The scheduler's task is to reorder a sequence of memory requests coming from memory processing units. After reordering, the new request sequence is forwarded to the external memory. The DRAM module permits significant performance gains if the reordered memory sequence increases the total DRAM row buffer hits. Since switching from one row in the row buffer to another achieves the worst access latency, reordering requests such that the incoming requests with the same row index are accessed sequentially is the objective of the memory scheduler. 

\begin{figure}[ht!]
\centering
\includegraphics[scale=0.25]{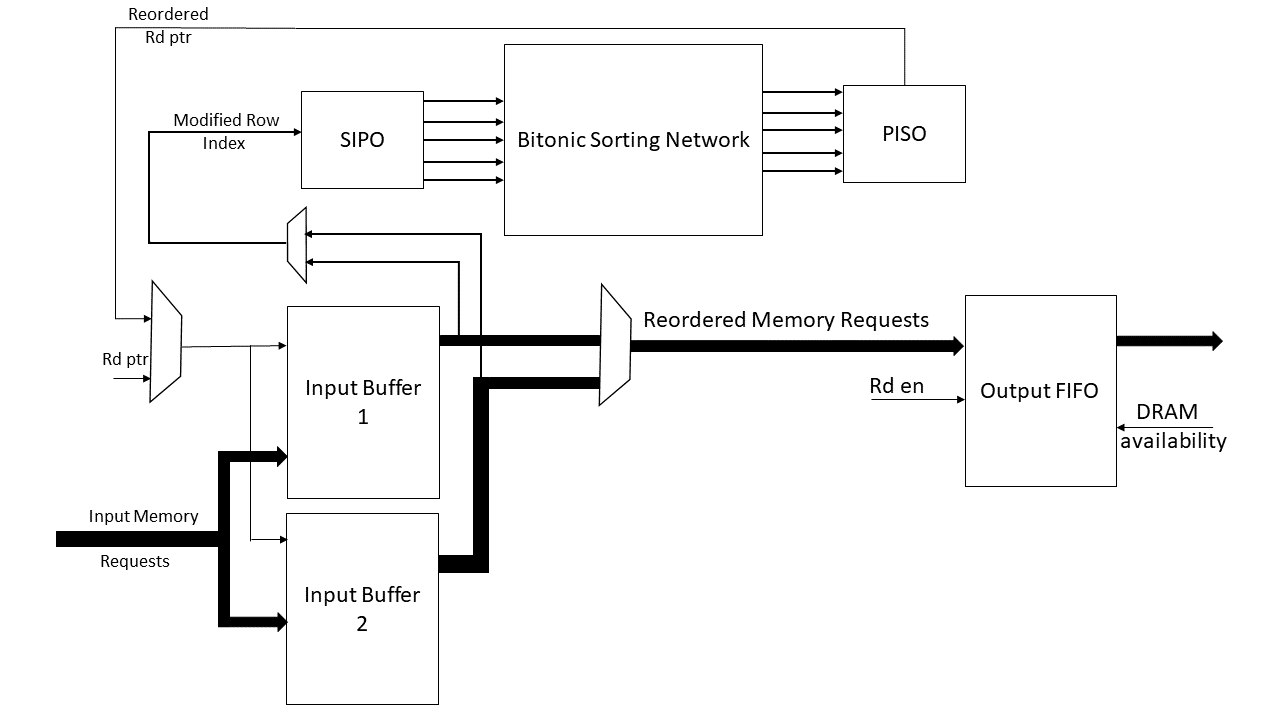}
\caption{Scheduler Architecture}
\label{Scheduler}
\end{figure}

All incoming requests are stored in input buffers as shown in Figure \ref{Scheduler}. Our current architecture uses double buffering to accommodate very high memory traffic. Each buffer stores the incoming requests depending on the destination DRAM bank. It also includes a timeout counter which increments every clock after storing the initial request to avoid deadlocks. Incoming requests to the buffers are appended with the current read pointer value. The batch formation is complete once any of the buffers become full or the timeout expires. A batch can only include one request type (read or write). It improves the overall performance of the scheduler and helps to maintain the weak consistency model our design maintain. Once a batch is formed, the row index from the address field and read pointer values are sent to a serial to parallel interface.  This stage involves data conditioning such that the row indexes can be parallelly sorted using a Bitonic sorting network. The sorting network takes $\frac{(\log{N})(\log{N} + 1)}{2}$ cycles to reorder the requests in the current batch based on the row address. This reordering process allows similar row indexes to be sequentially accessed. During the reordering process, the Bitonic sorting network leverages the parallelism enabled by FPGAs. The output of the sorting network feeds into a parallel to serial interface. The next step involves reading the reordered memory requests and extracting the buffer location index. These read addresses are then sent to the input buffers, where the complete memory request can be acquired from the buffer and then forwarded to the output FIFO. The sorted and batched requests can finally be forwarded to the DDR4 interface depending on the DRAM availability. 

The scheduler performance is quantified primarily by batch size and scheduler timeout. As described by R. Wei et al. \cite{electronics10040438}, addressing formats also invariably affect memory performance. 
\\
\\
\textbf{DRAM Timing Model:} 
An accurate DRAM timing model is the underpinning of a sound performance model. This can be utilized to calculate accurate latency improvements when using our scheduler. The primary DRAM timing parameters considered here are Column Address Strobe (CAS) latency ($T_{cl}$), Row Precharge Time ($T_{rp}$) and Row Address to Column Address Delay ($T_{rcd}$). We try to estimate memory access latency for two types of access patterns -  sequential and random. We also assume DRAM processes all the requests with \textit{open row policy}. For sequential accesses, the first-row buffer hit involves pre-charging a row and choosing a column. Therefore, the first row buffer hit is given by $T_{cl} + T_{rcd}$ delay. All subsequent row buffer hits have a $T_{cl}$ delay. Random memory requests have a first row buffer hit of $T_{cl} + T_{rcd}$ but each subsequent request would be a row conflict and such are given by $T_{rp} + T_{cl} + T_{rcd}$.
To create a better timing model for FPGA, we introduce average sequential latency ($T_{mem\_seq}$) and average random latency($T_{mem\_rand}$). $T_{mem\_seq} $ is given by
$\frac{T_{cl}\times T_{mem}}{T_{fpga}}$, where $T_{mem}$ is the clock period of a DRAM module and $T_{fpga}$ is the clock period of the FPGA. Similarly, $T_{mem\_rand}$ is given by $ \frac{(T_{rp} + T_{cl} + T_{rcd})\times T_{mem}}{T_{fpga}}$. As inferred from the DRAM timings, row index hits can have 2x - 3x the latency savings compared to row misses or row conflicts.
\\
\textbf{Memory Scheduling Time:} 
It is important to quantify the memory scheduling time to determine the performance of the memory scheduler. As Equation \ref{sch_eqn} indicates, for a batch of size N, we calculate two distinct values - first batch formation time and batch reordering time. The memory scheduling time mainly depends on the batch size. The $L_{data\_cond}$ in Equation \ref{sch_eqn} is the parallel to serial conversion latency and vice-versa ($\leq$ 2 cycles).
\begin{dmath}
{T}_{sch} = N + \frac{(\log{N})(\log{N} + 1)}{2} + L_{data\_cond}
\label{sch_eqn}
\end{dmath}

Our memory scheduler can be optimized for the FPGA-based applications favoring customizability and ease of reconfiguration as opposed to the raw speed that pure ASICs have. It also means being fully capable of utilizing the massively parallel computing power of an FPGA. The design adopts a hardware-accelerated parallel sorting network within the scheduler to do fast reordering. Sorting plays a significant role in the scheduler to allow the memory requests to be ordered in a manner most suitable to exploit row buffer hits. The second point of distinction is the use of parameters. Since FPGAs are inherently re-configurable, a simple script/parameter change allows the end-user to tailor the scheduler to its maximum efficiency. Since the scheduler only looks at its batch to do reordering, it allows further scalability. Like emphasized earlier, our work focuses on Modularity and programmability. The end-user only has to, therefore, change the batch size parameter in the HDL code without having to access or change the underlying hardware description to customize it.





The section \ref{sub_cache} and \ref{sub_dma} describe the cache unit and DMA Unit of our design.


\subsection{Cache Engine} \label{sub_cache}
The cache engine focuses on satisfying a single memory request with minimum latency. Within the cache engine, we explore the spatial and temporal locality while using a reconfigurable cache.

\begin{figure}[ht!]
\centering
\includegraphics[width=0.8\linewidth]{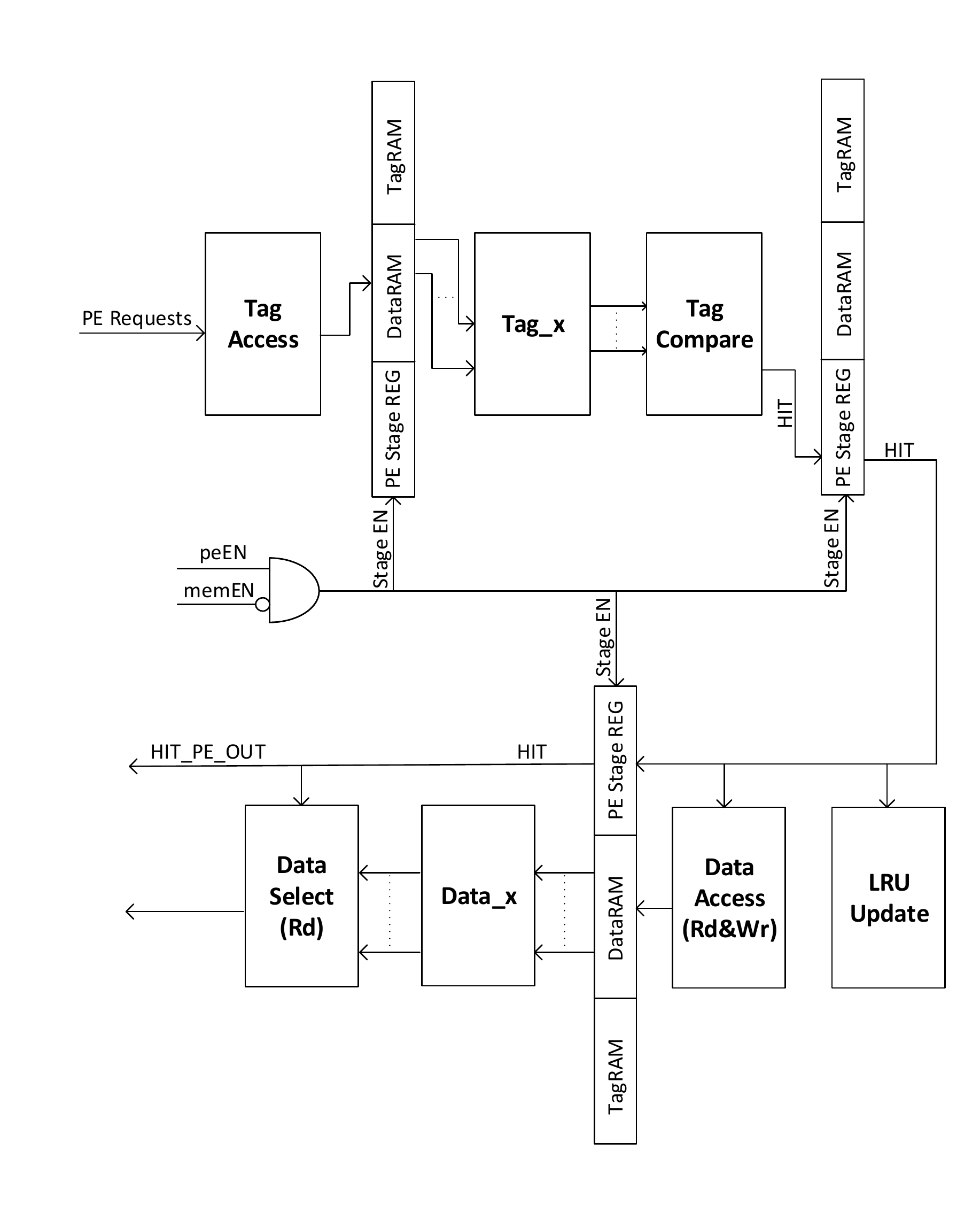}
\caption{Process element (PE) pipeline}
\label{pe_pipe}
\end{figure}
\begin{figure}
\centering
\includegraphics[width=0.9\linewidth]{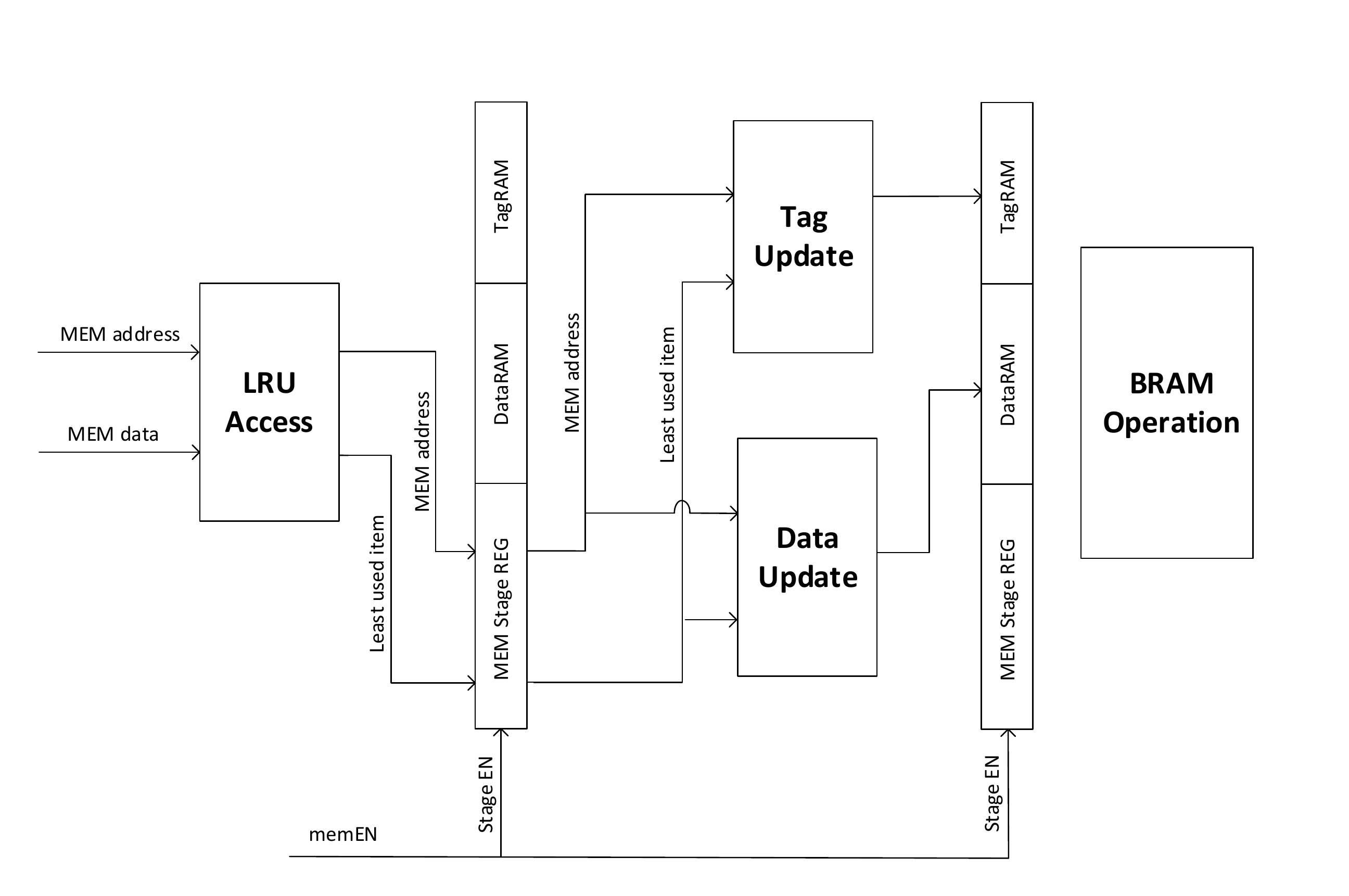}
\caption{Memory (MEM) pipeline}
\label{mem_pipe}
\end{figure}

\textbf{Cache:} 
Cache implementation uses two separate pipelines to achieve high frequency. Figure \ref{pe_pipe} and Figure \ref{mem_pipe} depict the PE pipeline and memory pipeline (MEM pipeline), respectively. These two pipelines share enable signals to indicate the requests sending from the PE or the memory are available to be processed. Only one pipeline is active based on the available data. While one pipeline is active, the other pipeline needs to be stalled because they share the same Tag RAM, Data RAM, and LRU. The PE pipeline is made into four stages, starting with a tag access step. Based on the address of the PE requests, tags are pulled out from the Tag RAM, denoted as \textit{Tag$\_$x}, and then compared to the incoming tag in the next stage. After the Tag comparator, cache hit information is generated and sent into the third stage. In this stage, the HIT information will be used as an evaluation criterion on whether the LRU update is needed or not. For read requests of \textit{m} (associativity) number of data, notated as \textit{Data$\_$x}, the data is pulled out from the Data RAM at the same time. Otherwise, for a write request with a hit, the updated data will be written into the corresponding entry of the Data RAM.

Slightly shorter than the PE pipeline, the MEM pipeline has 3 stages in total. The first stage for the memory pipeline is for the LRU lookup, where the \textit{location to be replaced} for the new incoming data is provided from the \textit{least recently used entry}. The next two stages are used for updating the corresponding locations in both Tag and Data RAM. The original tag and data placed at that location substitute as \textit{tag replace} and \textit{data replace}. Both of them are going to be returned to the DRAM if they get modified by a write request from PE.


The two pipelines share Tag RAM, Data RAM, and LRU. Therefore, an extra BRAM operation is used in the pipelines to avoid conflicts. It also ensures all the entries are up to date. Moreover, the memory pipeline operation has priority over the PE pipeline. Therefore, whenever a memory request is processed (i.e., indicates using high memory enable (memEN) signal), the processing pipeline is stalled.

Equation \ref{cache_eqn} shows the total time the cache unit takes when there are N consecutive accesses. We have assumed the DMA engine is not active to simplify the equation. $\mathds{1} (cache \; hit)$ indicates the current request is already in the cache.
 $\mathds{1}(cache \; miss \; active$ indicates the existence of cache misses in the cache.
\begin{dmath}
T_{cache} = L_{ctrl\_oh} \; + \; L_{cache}\;+ \\
\sum_{n=1}^{N} (\mathds{1}(cache \; hit \; \text{=} \; true) \\
(1 - \mathds{1}(active \; cache \;miss\; \text{=} \; true)) + \mathds{1}(cache \; miss\; \text{=} \; true)(L_{mem} + T_{sch} + T_{mem\_acc}))
\label{cache_eqn}
\end{dmath}

\subsection{DMA Engine} \label{sub_dma}
The DMA engine processes bulk transfers coming from the PEs. A DMA engine can have several DMA buffers inside. Therefore, it can support several bulk transfer requests from different PEs in parallel. The number of DMA buffers is a reconfigurable parameter, as discussed in section \ref{reconfig_mem}. A DMA request can contain one or more FLITs. When the first FLIT of a bulk transfer reaches the DMA engine, it passes onto one of the DMA buffer controllers. The corresponding DMA controller updates its status registers to \textit{occupied} and shares the PE ID of origin with the DMA Request Mapper. When another FLIT with the same PE ID comes to the memory controller, it routes to the same DMA buffer. DMA buffer controllers wait until all the FLITs of the bulk transfer are available before starting the external memory access. Equation \ref{eqn_dma} shows the total time takes to complete a DMA transfer with N elements assuming the cache engine is empty. $\mathds{1} (mem\_acc \; = \; seq)$ indicates the current access under consideration is a sequential access to the DRAM while $\mathds{1}(mem\_acc \; = \; rand)$ stands for a random access to the DRAM. A DMA request should either be random or sequential (i.e., $\mathds{1} (mem\_acc \; = \; seq) \; + \; \mathds{1} (mem\_acc \; = \; rand)\; = \; 1$). Additional latency of  ($L_{data\_convert}$) exists due to data width conversions. This is because typically, PE input/output data widths do not align with DRAM interface widths. 
\begin{multline}
T_{dma} = L_{ctrl\_oh} \; + \; T_{sch}\;+ L_{data\_convert}\; \\ + \sum_{n=1}^{N} (\mathds{1}(mem\_acc\; =\; seq)\times T_{mem\_seq} \\ + \mathds{1}(mem\_acc \;=\; rand)\times \; T_{mem\_rand})
\label{eqn_dma}
\end{multline}

The main advantages of having a DMA engine are: 
(a) DMA requests can request more than one element at once unlike the cache engine and reduces the input traffic of the memory controller,
(b) Using a DMA engine to access data with less spatial or temporal locality prevents cache pollution, and
(c) DMA transfers can utilize the memory bandwidth better for bulk transfers.
\\\\
\textbf{Memory Consistency Model:}
Our design follows a weak memory consistency model. This model is sufficient for existing FPGA accelerator designs used in different domains such as GCN and CNNs. The memory access conventions are as follows:

 \begin{itemize}
\item\textbf{ Consistency of Cache Engine:} The cache engine processes its requests on a first-in first-served basis. Therefore, weak consistency is maintained.

\item \textbf{Consistency of DMA Engine:} 
The DMA engine also processes its requests based on a first-in-first-out basis.

\item \textbf{Consistency between Cache Engine and DMA Engine: }
When Cache Engine and DMA Engine both have requests inside, the memory controller first processes all the cache requests that reach before the first DMA request. Then, all the DMA requests are satisfied. Finally, all the cache requests that arrive in between the DMA requests are processed. The requests generator logic in PEs should avoid conflicts between cache and DMA in such a scenario.

\item \textbf{Consistency of the scheduler: }
All the requests in a batch can be either read or write. Even though the requests to different addresses are re-ordered, the requests to the same address maintain the original order as they reach the memory controller.
 \end{itemize}

\section{Evaluation}
\subsection{Experimental Methodology} \label{big_data_workload}

We implemented different configurations of the memory controller on Xilinx Alveo U250 FPGA \cite{xilinxalveo} using Verilog HDL. Simulations, synthesis, and place-and-route implementations were performed using Xilinx Vivado Design Suite 2020.2. In our work, we focus on accessing a single external DRAM component efficiently. We use Xilinx memory interface IP\cite{xilinxddr} to connect our design with the external memory. For the Alveo U250 board, Xilinx provides Memory Interface IP with 31-bit address width and 512-bit data width (with ECC turned on).
\begin{table}[ht!] 
\centering
\caption{Resource Utilization of the cache}
\label{table_cache_util}
\begin{tabular}{|c|c|c|c|c|c|c|}\hline
\textbf{Cacheline} & \textbf{DoSA} & \textbf{Number of} & \textbf{LUT} & \textbf{FF} & \textbf{BRAM} & \textbf{URAM}
\\
\textbf{Width} & & \textbf{cachelines} & (\%) & (\%) & (\%) & (\%)
\\\hline\hline
512	&	1	&	512	&	0.01	&	0.01	&	0.3	&	0	\\\hline
512	&	1	&	1024	&	0.01	&	0.01	&	0.56	&	0	\\\hline
512	&	1	&	4096	&	0.65	&	0.64	&	0.06	&	0.63	\\\hline
512	&	2	&	2048	&	0.55	&	0.31	&	1.15	&	0	\\\hline
512	&	2	&	8192	&	1.87	&	1.24	&	0.24	&	1.25	\\\hline
1024	&	2	&	8192	&	1.99	&	1.28	&	0.24	&	2.34	\\\hline
2048	&	2	&	8192	&	2.22	&	1.37	&	0.24	&	4.53	\\\hline
4096	&	2	&	8192	&	2.46	&	1.55	&	0.24	&	8.91	\\\hline
512	&	4	&	4096	&	1.78	&	0.65	&	2.38	&	0	\\\hline
512	&	4	&	16384	&	7.49	&	2.4	&	0.93	&	2.5	\\\hline
512	&	8	&	8192	&	3.25	&	1.24	&	3.07	&	0	\\\hline
512	&	8	&	32768	&	9.79	&	3.1	&	2.1	&	5.4	\\\hline

\end{tabular}
\end{table}

The experiment section of this paper shows a detailed analysis of the performance and scalability of the proposed architecture. 
We use synthetic data, including weights, and inputs to create memory access patterns during the experiments. They are reflective of the memory access patterns of real-world datasets. We use generated datasets to experiment with Graph Convolution Networks (GCN) and Convolutional Neural Network (CNN).
\\
\textbf{GCN Memory Access Patterns:} Large scale FPGA based GCN accelerators\cite{9153263}\cite{10.1145/3373087.3375312}\cite{DBLP:journals/corr/HamiltonYL17} have 2 types of memory accesses: Feature Vector and Adjacent Vertex List. Feature vector access is a bulk transfer that typically has a size of 1 KB - 8 KB. Adjacent Vertex List consists of re-usable information that can be shared among PEs. We can exploit the data reuse of the Adjacent Vertex List by using the cache unit. The data reuse is approximately equal to $(\frac{Cache\;Size}{Size\;of\;the\;graph}) \times Average\;Degree$. The size of Adjacent Vertex List is 128 B – 512 B. 
\\
\textbf{CNN Memory Access Patterns:} CNN accelerators\cite{DBLP:journals/corr/SzegedyLJSRAEVR14}\cite{DBLP:journals/corr/HeZRS15}\cite{8445087} mainly have 2 types of memory accesses: layer kernel weights and layer inputs. The kernel size (each channel) of recent CNN models variate in the range of 1 x 1 - 11 x 11 (4 B – 512 B). These weights are accessed using the cache as they can be reused and are shared among PEs. The input access of each layer (per channel) varies between 1 KB – 16 KB.

\subsection{Resource Utilization}
\begin{figure}
\centering
\includegraphics[width=\linewidth]{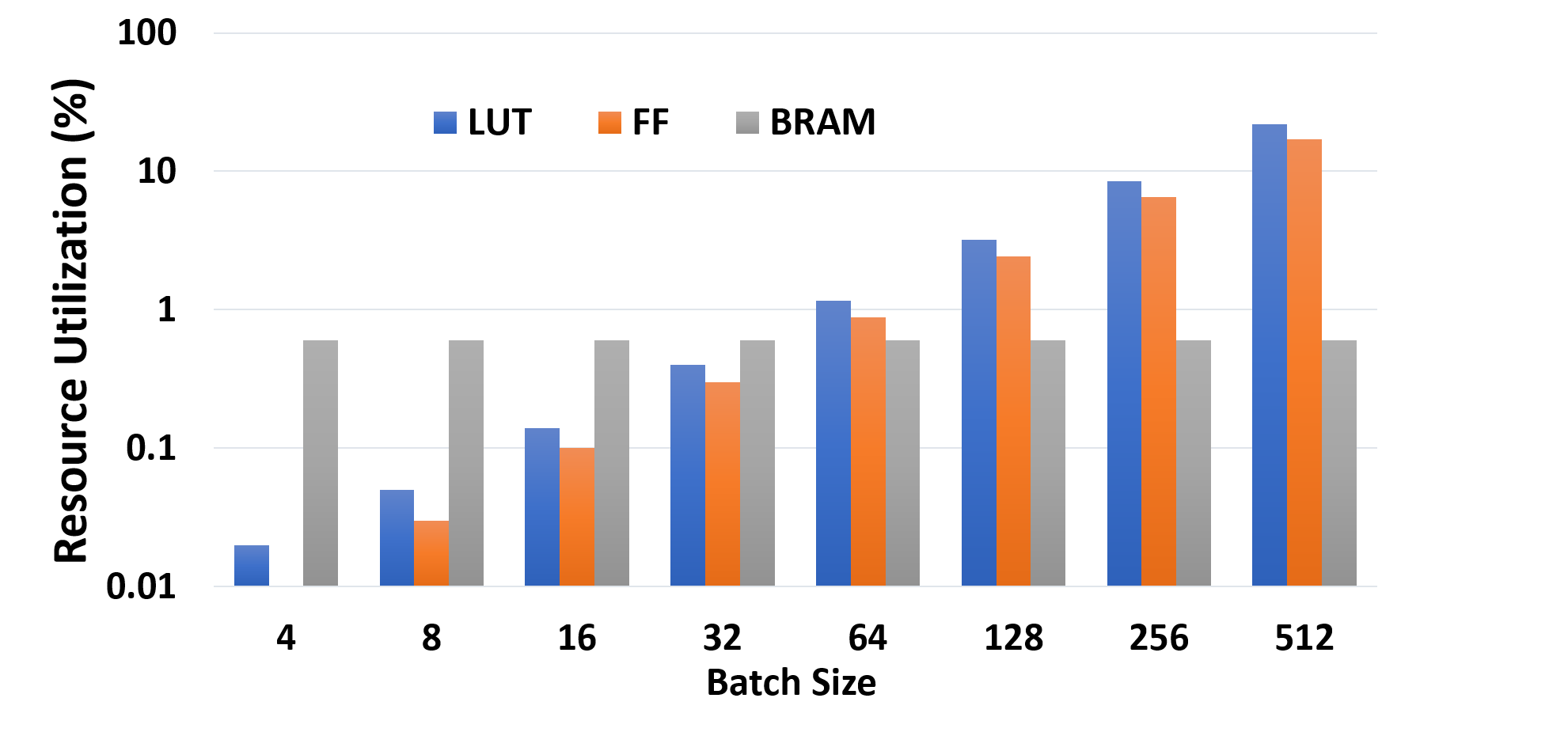}
\caption{Resource utilization of the scheduler for varying batch sizes}
\label{SizevsResource}
\end{figure}

This section focuses on the resource utilization of the main modules as we vary the reconfigurable parameters.
\\
\textbf{Cache:} Table \ref{table_cache_util} shows the resource consumption of the cache wrt. its reconfigurable parameters. In the cache implementation, data and tags are stored in URAMs. Therefore, URAM consumption linearly increases with the increase in degree of set-associativity (DoSA), cache line width, and the total number of cache lines.
\begin{figure}[ht!]
     \centering
     \begin{subfigure}[b]{0.8\linewidth}
         \centering
         \includegraphics[width=\textwidth]{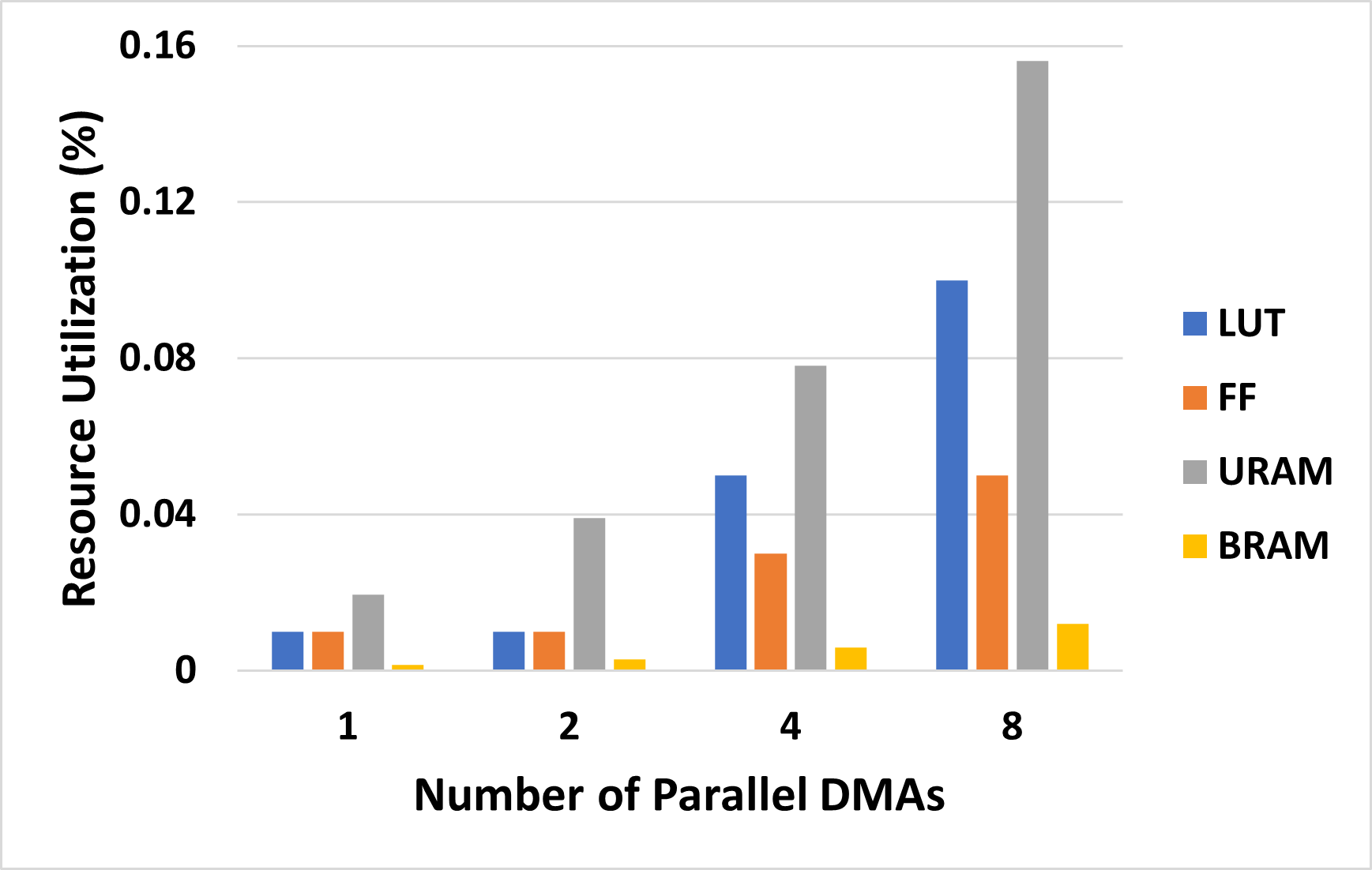}
         \caption{Size of DMA buffer = 8 KB}
         \label{dma_8kb}
     \end{subfigure}
     \hfill
     \begin{subfigure}[b]{0.8\linewidth}
         \centering
         \includegraphics[width=\textwidth]{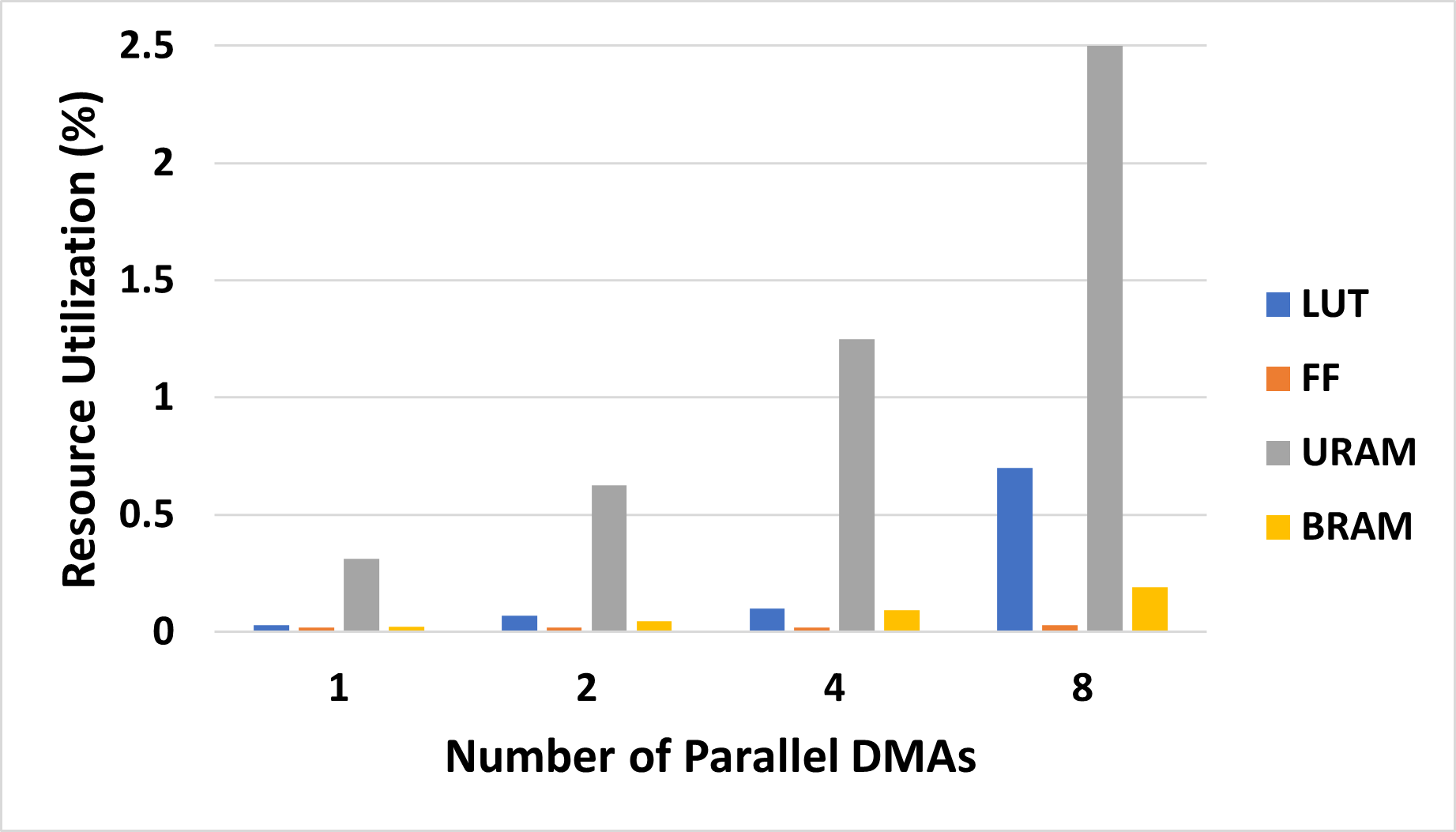}
         \caption{Size of DMA buffer = 128 KB}
         \label{dma_128kb}
     \end{subfigure}
     \hfill
     \begin{subfigure}[b]{0.8\linewidth}
         \centering
         \includegraphics[width=\textwidth]{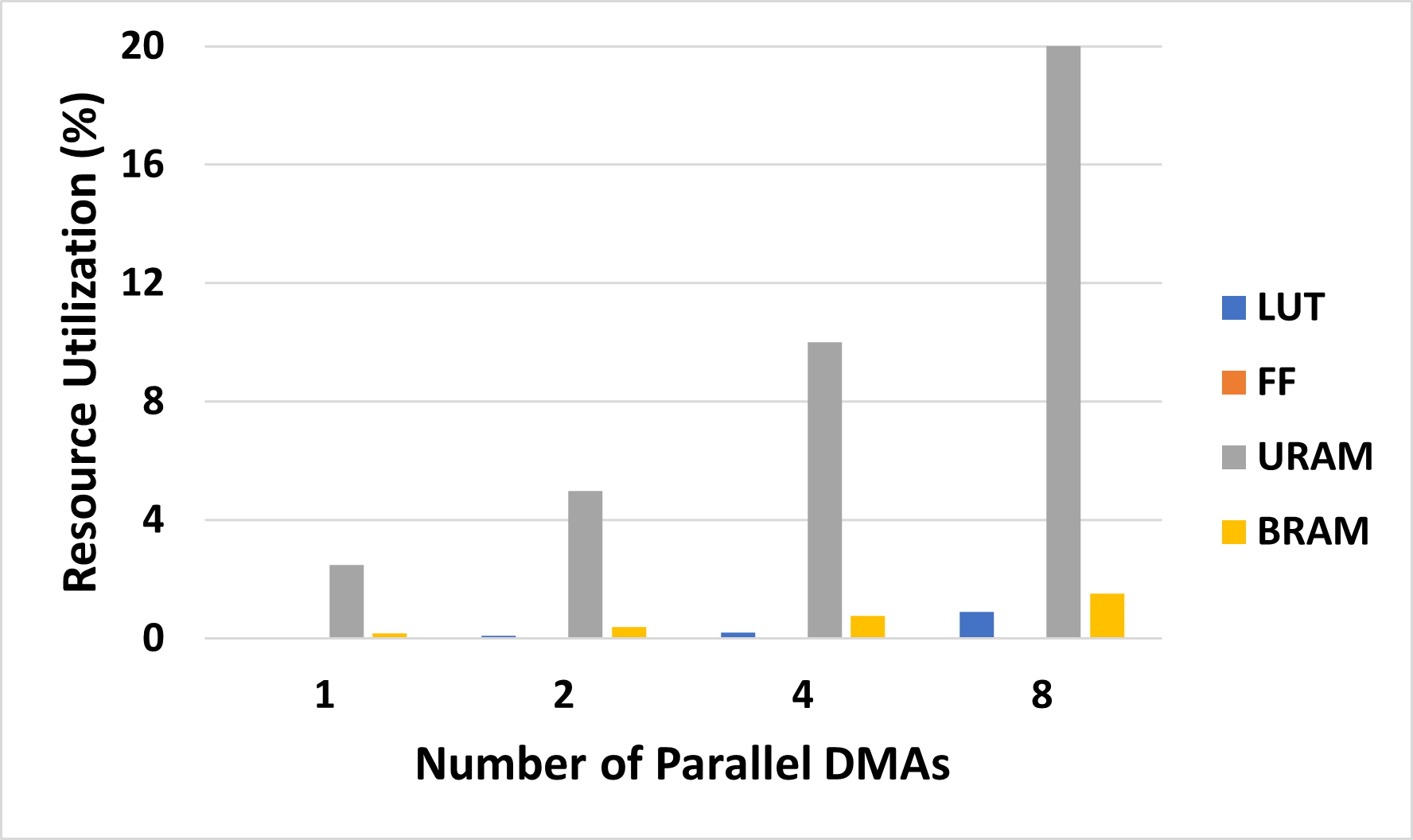}
         \caption{Size of DMA buffer = 1 MB}
         \label{dma_1mb}
     \end{subfigure}
        \caption{Resource utilization of DMA Engine}
        \label{resource_dma}
\end{figure}
\\
\textbf{DMA Unit:} Figure \ref{resource_dma} shows the resource utilization of the DMA Unit for different DMA buffer sizes and different parallel DMAs. We observe that URAM utilization linearly increases with the increase in simultaneous DMAs and DMA buffer size. Look-up Tables (LUT) and Flip FLop (FF) resource consumption stays below 2\% for all cases.
\\
\textbf{Scheduler:}
Figure \ref{SizevsResource} shows the resource utilization as batch sizes increase. BRAM utilization remains relatively constant at 0.60\% throughout all batch sizes. FF and LUT usage increase approximately by 3x every time batch sizes are doubled. Batch sizes above 512 become impractical to implement with scheduler utilization percentage alone reaching more than 20\% resource utilization in LUT and FF resources.

The rest of the modules consume less than 2\% of FPGA resources in all the experimented configurations.

\begin{table}[ht!] 
\centering
\caption{Module Parameters}
\label{our_configurations}
\begin{tabular}{|c|c|}\hline
\textbf{Module} & \textbf{Configuration}
\\\hline\hline
Cache	&	Cache line = 512 bit, DoSA = 4 \\ & Number of cache lines = 4096\\\hline
DMA Unit	& Buffer size = 16 KB \\ & Number of DMAs = 4 \\\hline
\end{tabular}
\end{table}
\subsection{Performance Analysis}

We configure our memory controller to support the common configurations of GCN as well as CNN (See Table \ref{our_configurations}) while conducting the experiments.

\textbf{Baseline:} We consider commercial\cite{xilinxddr}\cite{intelddr} memory interface IP directly connected to the accelerator as our baseline. Table \ref{our_configurations} shows the configurations we use to conduct following experiments.

\begin{figure}
     \centering
     \begin{subfigure}[b]{0.8\linewidth}
         \centering
         \includegraphics[width=\textwidth]{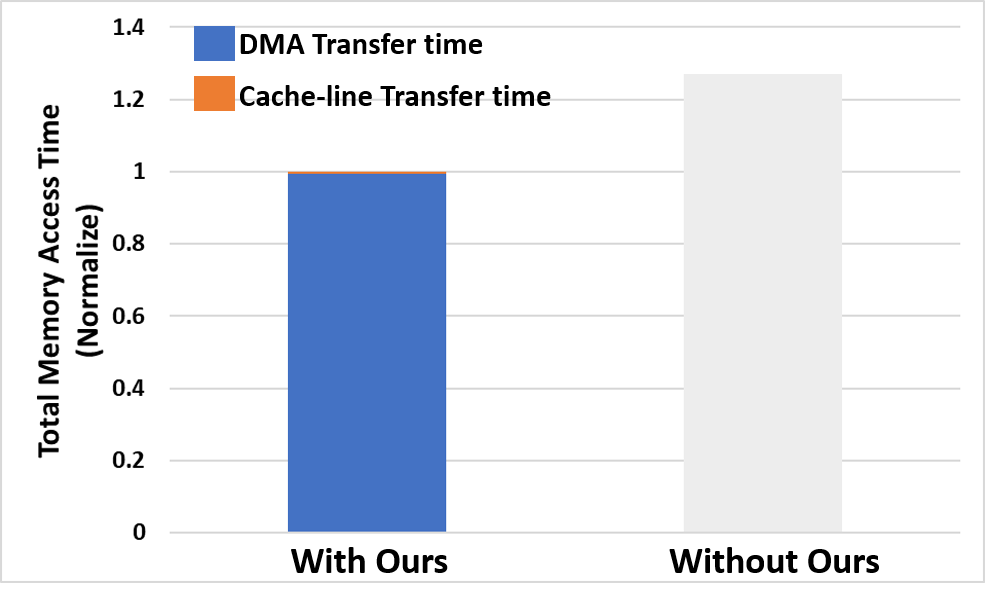}
         \caption{GCN inference time comparison}
         \label{gcn_workloads}
     \end{subfigure}
     \begin{subfigure}[b]{0.8\linewidth}
         \centering
         \includegraphics[width=\textwidth]{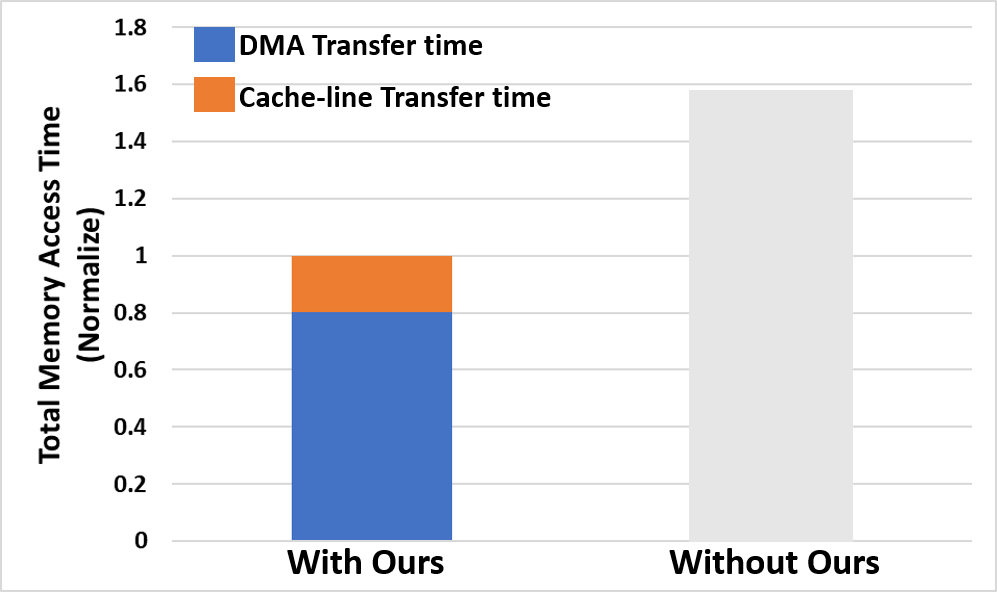}
         \caption{CNN inference time comparison}
         \label{cnn_workloads}
     \end{subfigure}
        \caption{Impact on time for different machine learning workloads}
        \label{workloads}
\end{figure}

Figure \ref{workloads} shows the memory access time comparison on different machine learning workloads between our implementation and the baseline. To clearly show the impact of our design and reduce the size of the graph, we divided the memory access time of each instance by the total time taken by our memory controller.

As shown in Figure \ref{gcn_workloads}, the total memory access time on GCN inference workloads are reduced by 27\% compared to the baseline. For this experiment, we used a synthetic graph with 1.6 M vertices and 240 M edges. Each vertex has 1024 features. We used the DMA engine to access the vertex features and the cache engine to access edges. The memory controller spent 99\% of 
time utilizing the DMA engine.

Figure \ref{cnn_workloads} shows the total memory access time on CNN inference workloads. Our memory controller has reduced memory access time by 58\% compared with the baseline. The experiments are conducted on the input layer of ResNet\cite{DBLP:journals/corr/HeZRS15} with 227 $\times$ 227 input images. We use the cache engine to access the image data while using the DMA engine to access weights. The results also indicate that our memory controller spent 80\% of the memory access time doing bulk transfers using the DMA engine.

\begin{figure}
\centering
\includegraphics[scale=0.43]{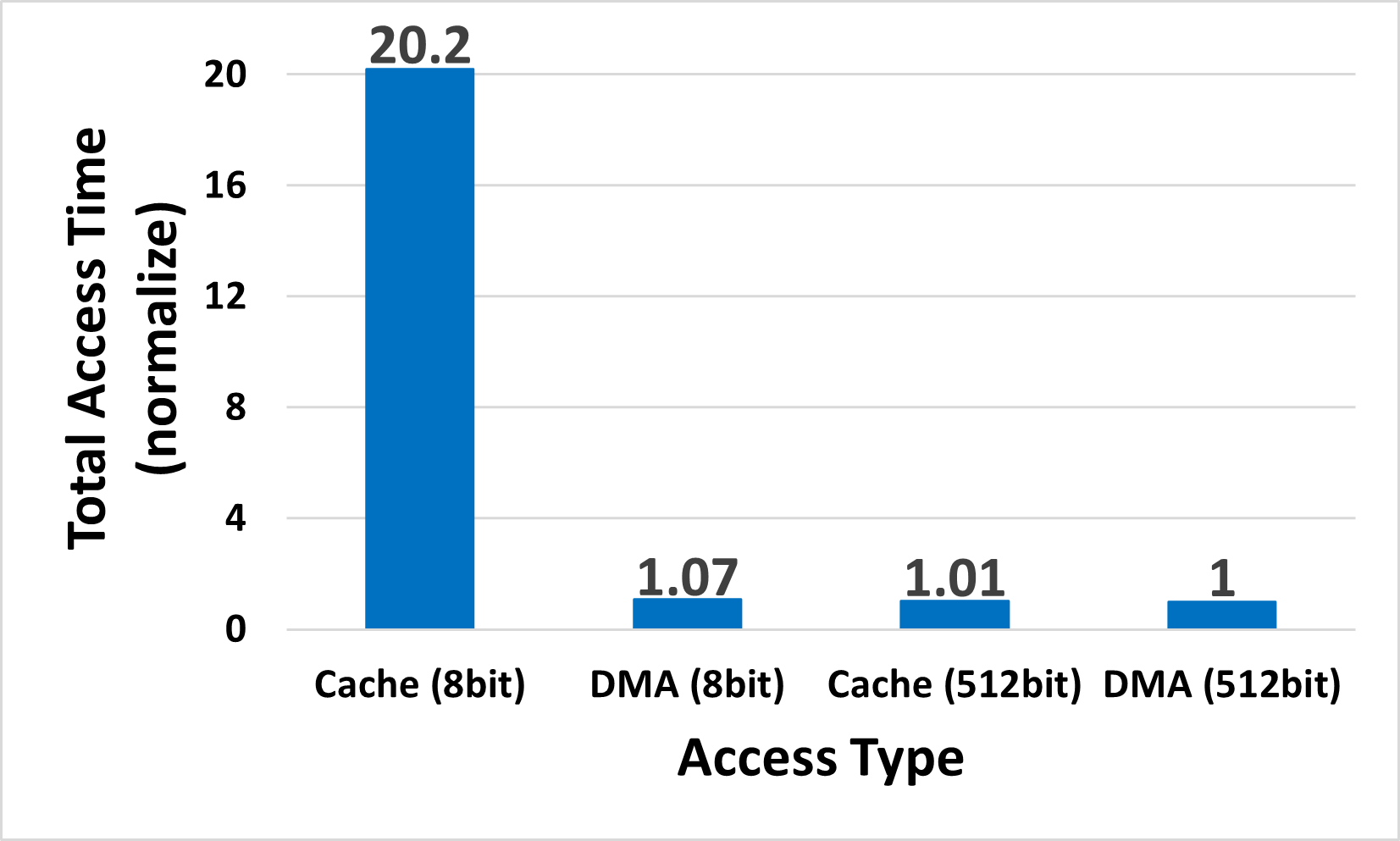}
\caption{Total access time for different PE to memory controller interface widths}
\label{acceess_time}
\end{figure}
Figure \ref{acceess_time} shows the total time to access 16 KB of sequential data with different interface widths between the memory controller and PEs. When small input widths are used, the cache underutilizes the memory bandwidth. Even though requests are sequential, cache miss latency on the first element of each line hurts the total access time. Unlike the cache, DMA can internally initiate bulk transfers to the memory. The external memory bandwidth can be used optimally with DMA transfers. Due to the aforementioned reasons, our system can obtain a 20x reduction in total access time compared to cache-only systems with narrow input interfaces.



Quantizing the performance of the scheduler involves two properties - Data Access patterns and batch size parameter. 
We utilize two common access patterns - Sequential access and random access. 
As the batch size increases, the probabilities of multiple requests with the same row index increases. The scheduler is more efficient, when the clock cycles saved from row index hits offsets the overall scheduling overhead. 
While the DRAM processes the first batch of requests, the second set of requests in the scheduler could be batched and ready to be sent to the DRAM. To further increase Scheduler efficiency, our scheduler also has the ability to completely bypass scheduling when the data access patterns are sequential or have low request traffic to avoid first batch latency for every batch.
\begin{figure}
\centering
\includegraphics[width=0.8\linewidth]{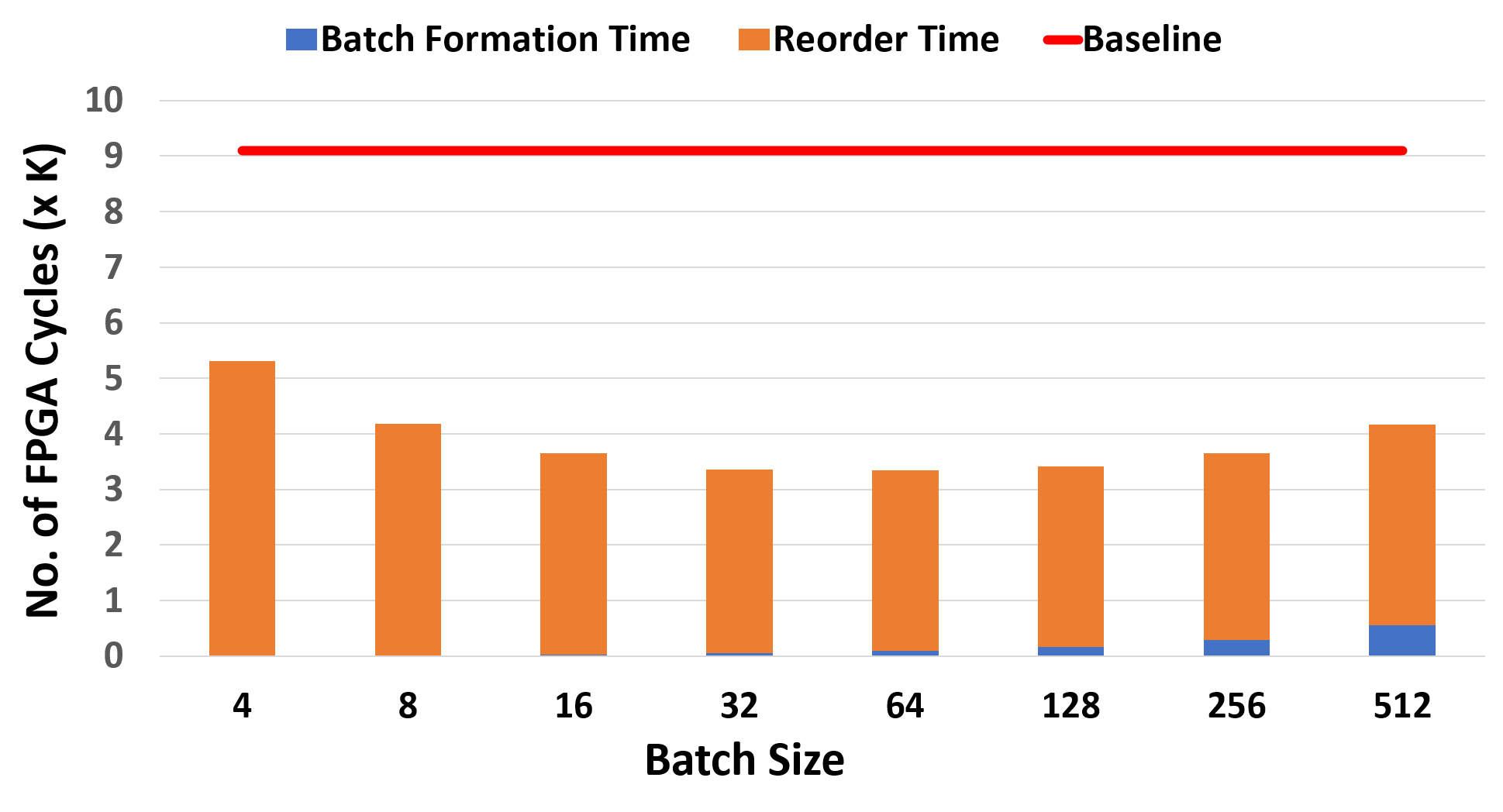}
\caption{Schedule time breakdown for varying batch sizes}
\label{SizevsSpeed}
\end{figure}

We use data access patterns which are mentioned in Section \ref{big_data_workload}. 
The batch formation time is the total amount of time the initial batch needs to be fully processed by the scheduler. As shown in Figure \ref{SizevsSpeed}, batch formation time increases with an increase in batch size even when the time complexity of our Bitonic sorting network grows slower than the batch formation time $N$.  This is because every subsequent batch formation latency can be overlapped with the DRAM processing time. Consequently, the second metric describes the total time it takes to complete batching and processing of the batches by the DRAM. It is clearly seen that increasing the Batch size leads to a decrease in overall latency until the point where scheduler overhead causes performance deterioration. We find that batch sizes of 32 and 64 provide the highest performance while maintaining modest resource utilization. 



\section{Conclusion}
In this paper, we proposed a programmable DRAM memory controller 
to use with custom FPGA accelerators. Our modular hardware consists of reconfigurable core components: cache, DMA, and memory scheduler. The programmable memory controller allows the end-user to optimize external memory access for an application of choice while allowing the utilization of the remaining hardware resources after implementing the targeted application design on FPGA. 
In the future, we will explore FPGA-based heterogeneous memory platforms and further optimize our memory controller towards such platforms.

\section{Acknowledgment}
This work was supported by the U.S. National Science Foundation (NSF) under grant OAC-1911229 and in part by Xilinx.

\bibliographystyle{IEEEtran}
\bibliography{hpec20}

\end{document}